\documentclass[floats,preprint,nofootinbib,superscriptaddress,tightenlines]{revtex4}

\usepackage{amsfonts,amsmath,amssymb,slashed}
\usepackage{bm}
\usepackage[colorlinks,anchorcolor=black,citecolor=violet,linkcolor=blue]{hyperref}
\usepackage{graphicx}
\usepackage{multirow}
\usepackage{xcolor}

\begin{document}
\baselineskip=17pt \parskip=5pt

\preprint{NCTS-PH/2008}
\hspace*{\fill}

\title{Kaon decays shedding light on massless dark photons}

\author{Jhih-Ying Su}
\affiliation{Department of Physics, National Taiwan University, Taipei 106, Taiwan}

\author{Jusak Tandean}
\affiliation{Department of Physics, National Taiwan University, Taipei 106, Taiwan}
\affiliation{Physics Division, National Center for Theoretical Sciences, Hsinchu 300, Taiwan
\bigskip}


\begin{abstract}

We explore kaon decays with missing energy carried away by a massless dark photon,
$\overline\gamma$, assumed to have flavor-changing dipole-type couplings to the $d$
and $s$ quarks.
We consider in particular the neutral-kaon modes $K_L\to\gamma\overline\gamma$ and
$K_L\to\pi^0\gamma\overline\gamma$ and their $K_S$ counterparts, as well as the charged-kaon
channel $K^+\to\pi^+\gamma\overline\gamma$, each of which also has an ordinary photon, $\gamma$,
in the final state.
In addition, we look at $K_{L,S}\to\pi^+\pi^-\overline\gamma$ and
$K^+\to\pi^+\pi^0\overline\gamma$.
Interestingly, the same $ds\overline\gamma$ interactions give rise to the flavor-changing
two-body decays of hyperons with missing energy and are subject to
model-independent constraints that can be inferred from the existing hyperon data.
Taking this into account, we obtain branching fractions ${\cal B}(K_L\to\gamma\overline\gamma)$
and ${\cal B}(K_L\to\pi^0\gamma\overline\gamma)$ which can be as high as $10^{-3}$ and
$10^{-6}$, respectively, one or both of which may be within the sensitivity reach of
the KOTO experiment.
Furthermore, we find that ${\cal B}(K^+\to\pi^+\gamma\overline\gamma)$ and
${\cal B}(K^+\to\pi^+\pi^0\overline\gamma)$ are allowed to be maximally of order
$10^{-6}$ as well, which may be probed by NA62.
Complementarily, the hyperon modes can have rates which are potentially accessible
by BESIII.
Thus, these ongoing experiments could soon be able to offer significant tests on
the existence of the massless dark photon.

\end{abstract}

\maketitle

\section{Introduction\label{intro}}

Over the past few decades various phenomenological considerations have motivated the introduction
of the so-called dark photon, a spin-one boson associated with a new Abelian gauge symmetry,
U(1)$_D$, under which all the fields of the standard model (SM) are singlets~\cite{Jaeckel:2010ni,
Essig:2013lka,Alexander:2016aln,Beacham:2019nyx,Gninenko:2020hbd,Fabbrichesi:2020wbt,Holdom:1985ag,
delAguila:1995rb,Hoffmann:1987et,Dobrescu:2004wz,Gabrielli:2016cut,Fargion:2005ep,Fabbrichesi:2017vma,
Fabbrichesi:2019bmo,Su:2019ipw,Su:2020yze,Zhang:2018fbm,Foot:2014uba,Foot:2014osa,Foot:1991kb,
He:2017zzr,Fayet:2006sp,Pospelov:2008zw,Reece:2009un,Chiang:2016cyf}.
The dark photon may be massive or massless, depending on whether U(1)$_D$ is spontaneously broken
or stays unbroken, respectively.
The massive one, often symbolized by $A'$, can interact directly with SM fermions through
a renormalizable operator, $\epsilon eA_\mu'J_{\textsc{em}}^\mu$, which involves the electromagnetic
current $eJ_{\textsc{em}}$ and a small parameter $\epsilon$ due to the kinetic mixing between
the dark and SM Abelian gauge fields~\cite{Jaeckel:2010ni,Essig:2013lka,Alexander:2016aln,
Beacham:2019nyx,Gninenko:2020hbd,Fabbrichesi:2020wbt,Holdom:1985ag}.
It follows that $A'$ could be produced in the decays or scatterings of SM fermions and hadrons
and it might decay into electrically charged fermions or mesons.
In general, it could also decay invisibly into other dark particles.
These possibilities have stimulated numerous dedicated quests for it, but with negative results
so far, leading to bounds on $\epsilon$ over various ranges of the $A'$ mass~\cite{Jaeckel:2010ni,
Essig:2013lka,Alexander:2016aln,Beacham:2019nyx,Gninenko:2020hbd,Fabbrichesi:2020wbt,
Batley:2015lha,Aaij:2017rft,Anastasi:2018azp,Ablikim:2018bhf,CortinaGil:2019nuo,NA64:2019imj}

The massless dark photon, here denoted by $\overline\gamma$, is very dissimilar from the massive
one because they differ substantially in both theoretical underpinnings and experimental
signatures~\cite{Fabbrichesi:2020wbt,Holdom:1985ag,delAguila:1995rb,Hoffmann:1987et,
Dobrescu:2004wz,Fargion:2005ep,Gabrielli:2016cut,Fabbrichesi:2017vma,Fabbrichesi:2019bmo,
Su:2019ipw,Su:2020yze,Zhang:2018fbm,Foot:2014uba,Foot:2014osa}.
If U(1)$_D$ remains unbroken, one can always arrange a linear combination of the dark and SM U(1)
gauge bosons such that it has no renormalizable connection to the SM and can then be identified
as the massless dark photon~\cite{Holdom:1985ag,Dobrescu:2004wz}.
Since it therefore does not interact directly with SM members, the limitations implied by
the aforementioned hunts for $A'$ are not applicable to~$\overline\gamma$.
Nevertheless, the latter could still have consequential impact via higher-dimensional
operators~\cite{Hoffmann:1987et,Dobrescu:2004wz,Gabrielli:2016cut}, caused by loop diagrams
containing new heavy particles, which may translate into detectable effects.
This suggests that potentially promising avenues to seek $\overline\gamma$ may be available
and hence should be explored.
Some of them will be put forward below, which may be feasible at ongoing or near-future experiments.
Given that the viable parameter space of the massive dark photon continues to shrink with
accumulating null outcomes of its searches, it is of great interest to pay increasing attention to
the alternate possibility that the dark photon is massless.

In this paper we concern ourselves with flavor-changing neutral current (FCNC) transitions induced
by the massless dark photon, $\overline\gamma$, having nonrenormalizable interactions with the $d$
and $s$ quarks described by dimension-five operators in the Lagrangian
\begin{align} \label{Ldsg}
{\cal L}_{ds\bar\gamma}^{} & \,=\, -\overline d \big( {\mathbb C}
+ \gamma_5^{} {\mathbb C}_5^{} \big) \sigma^{\mu\nu} s\, \bar F_{\mu\nu}^{}
\,+\, {\rm H.c.} \,, ~~~ ~~~~
\end{align}
where $\mathbb C$ and ${\mathbb C}_5$ are constants which have the dimension of inverse mass and
can be complex, \,$\bar F_{\mu\nu}=\partial_\mu\bar A_\nu-\partial_\nu\bar A_\mu$\, is the
field-strength tensor of $\overline\gamma$, and \,$\sigma^{\mu\nu}=i[\gamma^\mu,\gamma^\nu]/2$.\,
In the absence of other particles beyond the SM lighter than the electroweak scale,
${\cal L}_{ds\bar\gamma}$ could originate from dimension-six operators which respect
the SM gauge group and the unbroken U(1)$_D$.
One can express such operators in the form
\,${\cal L}_{\textsc{np}} = -\Lambda_{\textsc{np}}^{-2} \big( {\cal C}_{12}
\overline{\textsl{\texttt Q}_1} \sigma^{\mu\nu} d_2 + {\cal C}_{21}
\overline{\textsl{\texttt Q}_2} \sigma^{\mu\nu} d_1 \big) H \bar F_{\mu\nu} + {\rm H.c.}$,\,
where $\Lambda_{\textsc{np}}$ represents an effective heavy mass scale,
the dimensionless coefficients ${\cal C}_{12,21}$ are generally complex,
$\textsl{\texttt Q}_{1,2}$ ($d_{1,2}$) stand for left-handed quark doublets (right-handed
down-type quark singlets) from the first two families, and $H$ designates the SM Higgs
doublet~\cite{Dobrescu:2004wz}.
Accordingly
\,$\mbox{\small$\mathbb C$} \Lambda_{\textsc{np}}^2 =
\big({\cal C}_{12}^{}+{\cal C}_{21}^*\big)v/\sqrt8$\,
and
\,$\mbox{\small$\mathbb C$}_5^{} \Lambda_{\textsc{np}}^2 =
\big({\cal C}_{12}^{}-{\cal C}_{21}^*\big)v/\sqrt8$,\,
with $v\simeq246$\,\,GeV being the Higgs vacuum expectation value.
Both $\Lambda_{\textsc{np}}$ and ${\cal C}_{12,21}$ depend on the details of the underlying
new physics (NP).

The interactions in ${\cal L}_{ds\bar\gamma}$ bring about the FCNC decays of hyperons into
a lighter baryon plus missing energy carried away by the massless dark photon.
In Ref.\,\cite{Su:2019ipw} we have studied such two-body processes and demonstrated that
their rates are allowed by present constraints to reach values that are within
the sensitivity reach of the ongoing BESIII experiment~\cite{Li:2016tlt,Ablikim:2019hff}.
Analogous transitions can take place in the kaon sector.
In the case of massive dark photon, \,$K\to\pi A'$\, and \,$K^+\to\ell^+\nu A'$\, with
\,$\ell=e,\mu$\, might be useful in the quests for it \cite{Fayet:2006sp,Pospelov:2008zw,
Reece:2009un,Chiang:2016cyf,Batley:2015lha}.
In contrast, since $\overline\gamma$ is massless and has no renormalizable links to the SM,
angular-momentum conservation and gauge invariance forbid \,$K\to\pi\overline\gamma$,\, while
\,$K^+\to\ell^+\nu\overline\gamma$\, would be highly suppressed.
Instead, it has been suggested in Ref.\,\cite{Fabbrichesi:2017vma} that ${\cal L}_{ds\bar\gamma}$
could be probed with \,$K^+\to\pi^+\pi^0\overline\gamma$,\,
which might be accessible in the NA62 experiment~\cite{NA62:2017rwk}.

It turns out that there are other kaon modes which may provide additional and competitive
windows into the same $ds\overline\gamma$ couplings.
Specifically, here we propose to pursue the neutral-kaon channels \,$K_L\to\gamma\overline\gamma$\,
and \,$K_L\to\pi^0\gamma\overline\gamma$\, and the charged one
\,$K^+\to\pi^+\gamma\overline\gamma$,\, all of which have an ordinary photon, $\gamma$,
among the daughter particles.
As we will show, the two $K_L$ modes could have rates which may be big enough to
be observable in the currently running KOTO experiment~\cite{Ahn:2018mvc}.
We will also examine \,$K_S\to\gamma\overline\gamma,\pi^0\gamma\overline\gamma$\, and
\,$K_{L,S}\to\pi^+\pi^-\overline\gamma$\, and take another look at
\,$K^+\to\pi^+\pi^0\overline\gamma$.\,

The remainder of the paper is organized as follows.
In Sec.\,\,\ref{krates} we first deal with the amplitudes for the kaon decays being analyzed
and then calculate their rates.
In treating the amplitudes, we need the relevant mesonic matrix-elements of the quark bilinears
in Eq.\,(\ref{Ldsg}).
To derive them, we utilize the techniques of chiral perturbation theory.
In Sec.\,\,\ref{predict} we evaluate the maximal branching fractions of the kaon modes,
taking into account model-independent restrictions on the $ds\overline\gamma$ couplings
deduced from the available hyperon data.
We draw our conclusions in Sec.\,\,\ref{concl}.

\section{Kaon decay amplitudes and rates\label{krates}}

To investigate the influence of ${\cal L}_{ds\bar\gamma}$ on our processes of interest,
we adopt the framework of chiral perturbation theory~\cite{Gasser:1984gg}.
In this context, one can obtain the correspondences between operators comprising bilinears of
the quark fields \,$(\textit{\textsf q}_1,\textit{\textsf q}_2,\textit{\textsf q}_3)=(u,d,s)$\,
and their hadronic counterparts involving the lightest pseudoscalar-meson fields,
which constitute a flavor-SU(3) octet and are collected into
\begin{align}
\Sigma & \,=\, e^{i\varphi/f} \,, &
\varphi & \,= \sqrt2 \left( \begin{array}{ccc} \frac{1^{\vphantom{|}}}{\sqrt{2}}\,
\pi^0+\frac{1}{\sqrt6}\, \eta_8^{} & \pi^+ &  K^+ \medskip \\
\pi^- & \frac{-1}{\sqrt2}\, \pi^0 + \frac{1}{\sqrt6}\, \eta_8^{} & K^0 \medskip \\
K^- & \,\overline{\!K}{}^0 & \frac{-2}{\sqrt6}\, \eta_8^{} \end{array} \right) ,
\end{align}
where $f$ denotes the meson decay constant.
At the leading chiral order, the bosonization of the quark tensor currents in Eq.\,(\ref{Ldsg})
has been addressed before~\cite{Colangelo:1999kr,
Tandean:2000qk,Gao:2002ub,Mertens:2011ts,Kamenik:2011vy}.
Explicitly, it is most generally given by~\cite{Mertens:2011ts}
\begin{align} \label{qsq}
\overline{\textit{\textsf q}_{\textsl{\texttt I}}^{}}\sigma_{\mu\nu}^{} \big(1-\gamma_5^{}\big)
\textit{\textsf q}_{\textsl{\texttt J}}^{}\, &
\Leftrightarrow\; -i a_T^{} f^2 \Big[ \Big( {\cal D}_\mu^{}\Sigma
{\cal D}_\nu^{}\Sigma^\dagger - {\cal D}_\nu^{}\Sigma {\cal D}_\mu^{}\Sigma^\dagger
+ i\epsilon_{\mu\nu\varrho\varsigma}^{} {\cal D}^\varrho\Sigma {\cal D}^\varsigma
\Sigma^\dagger \Big) \Sigma \Big]_{\textsl{\texttt{JI}}}
\nonumber \\ & ~~~~ +\,
a_T' f^2 \Big[ \Sigma \Big({\texttt F}_{\mu\nu}^R+i\widetilde{\texttt F}{}_{\mu\nu}^R\Big)
+ \Big( {\texttt F}_{\mu\nu}^L + i\widetilde{\texttt F}{}_{\mu\nu}^L \Big)\Sigma
\Big]_{\textsl{\texttt{JI}}} \,,
\nonumber \\
\overline{\textit{\textsf q}_{\textsl{\texttt I}}^{}}\sigma_{\mu\nu}^{} \big(1+\gamma_5^{}\big)
\textit{\textsf q}_{\textsl{\texttt J}}^{}\, &
\Leftrightarrow\; -i a_T^{\vphantom{|_|^|}} f^2 \Big[ \Big(
{\cal D}_\mu^{}\Sigma^\dagger {\cal D}_\nu^{}\Sigma - {\cal D}_\nu^{}\Sigma^\dagger
{\cal D}_\mu^{}\Sigma - i\epsilon_{\mu\nu\varrho\varsigma}^{} {\cal D}^\varrho
\Sigma^\dagger {\cal D}^\varsigma\Sigma \Big) \Sigma^\dagger \Big]_{\textsl{\texttt{JI}}}
\nonumber \\ & ~~~~ +\,
a_T' f^2 \Big[ \Sigma^\dagger \Big( {\texttt F}_{\mu\nu}^L
-i\widetilde{\texttt F}{}_{\mu\nu}^L \Big) + \Big( {\texttt F}_{\mu\nu}^R
- i\widetilde{\texttt F}{}_{\mu\nu}^R \Big) \Sigma^\dagger \Big]_{\textsl{\texttt{JI}}} \,,
\end{align}
where $a_T^{}$ and $a_T'$ are constants having the dimension of inverse mass and electromagnetic
effects are included via\footnote{Under chiral
SU(3)$_L\times{\rm SU}(3)_R$ transformations \,$\Sigma\to V_L^{}\Sigma V_R^\dagger$,
\,${\cal D}_\mu\Sigma\to V_L^{}{\cal D}_\mu\Sigma V_R^\dagger$,\,
\,${\texttt F}_{\mu\nu}^\chi\to V_\chi^{}{\texttt F}_{\mu\nu}^\chi V_\chi^\dagger$,\, and
\,$\widetilde{\texttt F}{}_{\mu\nu}^\chi\to V_\chi^{}\widetilde{\texttt F}{}_{\mu\nu}^\chi
V_\chi^\dagger$\,
for \,$\chi=L,R$,\, where \,$V_\chi\in{\rm SU}(3)_\chi$.}
\begin{align}
{\cal D}_\mu^{}\Sigma & \,=\, \partial_\mu^{}\Sigma - i {\texttt F}_\mu^L \Sigma
+ i \Sigma\, {\texttt F}_\mu^R \,, &
{\texttt F}_\mu^L & \,=\, {\texttt F}_\mu^R \,=\, -e A_\mu^{} Q_q^{} \,,
\nonumber \\
{\texttt F}_{\mu\nu}^L & \,=\, {\texttt F}_{\mu\nu}^R \,=\, -eF_{\mu\nu}^{} Q_q^{} \,, &
\widetilde{\texttt F}{}_{\mu\nu}^L & \,=\, \widetilde{\texttt F}{}_{\mu\nu}^R \,=\,
-e\, \epsilon_{\mu\nu\varrho\varsigma}^{}\, \partial^\varrho A^\varsigma\, Q_q^{} \,, ~~
\nonumber \\
F_{\mu\nu}^{} & \,=\, \partial_\mu^{}A_\nu^{}-\partial_\nu^{}A_\mu^{} \,, &
Q_q^{} & \,=\, \tfrac{1}{3}\, {\rm diag}(2,-1,-1) \,,
\end{align}
with $A_\mu$ and $F_{\mu\nu}$ standing for the ordinary photon field and its field-strength tensor,
respectively, and $Q_q$\, representing the electric-charge matrix of the three lightest quarks.
Hence the subscript pair \,$\textsl{\texttt{JI}}=32\;(23)$\, on the right-hand sides of
Eq.\,(\ref{qsq}) corresponds to \,$s\to d$ ($d\to s$) transitions.

This allows us to determine the matrix elements required to write down the amplitudes for
\,$K\to\gamma\overline\gamma$\, and \,$K\to\pi\gamma\overline\gamma$\, arising from
${\cal L}_{ds\bar\gamma}$ which have both an ordinary photon, $\gamma$, and a massless dark
photon in the final states.
Thus, for \,$K\to\gamma\overline\gamma$\, we arrive at
\begin{align}
\big\langle\gamma\big|\overline d\sigma_{\alpha\omega}^{}s\big|\,\overline{\!K}{}^0\big\rangle
& \,=\, \big\langle\gamma\big|\overline s\sigma_{\alpha\omega}^{}d\big|K^0\big\rangle \,=\,
\frac{i\sqrt8}{3}\, a_T'ef\, \epsilon_{\alpha\omega\mu\nu}^{}\,\varepsilon^{*\mu\,}
{\texttt k}^\nu \,,
\nonumber \\ \label{<g|dsg5s|K>}
\big\langle\gamma\big|\overline d\sigma_{\alpha\omega}^{}\gamma_5^{}s
\big|\,\overline{\!K}{}^0\big\rangle
& \,=\, \big\langle\gamma\big|\overline s\sigma_{\alpha\omega}^{}\gamma_5^{}d\big|K^0\big\rangle
\,=\, \frac{\sqrt8}{3}\, a_T'e f\, \big( \varepsilon_\omega^* {\texttt k}_\alpha^{}
- \varepsilon_\alpha^* {\texttt k}_\omega^{} \big) \,,
\end{align}
where $\varepsilon$ and $\texttt k$ are the ordinary photon's polarization vector and
momentum, respectively.
Contracting these matrix elements with the dark photon's polarization vector $\bar\varepsilon$ and
momentum $\bar q$ as dictated by Eq.\,(\ref{Ldsg}), in conjunction with applying the approximation
\,$\sqrt2\,K_L=K^0+\,\overline{\!K}{}^0$,\, then yields the amplitude
\begin{align} \label{MK2gg'}
{\cal M}_{K_L\to\gamma\overline\gamma}^{} & \,=\, \frac{4 a_T' e f}{3} \bigl[
-\epsilon^{\mu\nu\varrho\varsigma}\, {\rm Re}\,\mbox{\small$\mathbb C$} + \big( g^{\mu\varsigma}
g^{\nu\varrho} - g^{\mu\nu} g^{\varrho\varsigma} \big)\, {\rm Im}\,{\mathbb C}_5 \big]
\varepsilon_\mu^*\bar\varepsilon_\nu^*\, {\texttt k}_\varrho^{} \bar q_\varsigma^{} \,.
\end{align}
This leads to the decay rate
\begin{align} \label{GK2gg'}
\Gamma_{K_L\to\gamma\overline\gamma}^{} & \,=\, \frac{8\alpha_{\rm e}^{}}{9} (a_T'f)^2
m_{K^0}^3 \big(|{\rm Re}\,{\mathbb C}|^2+|{\rm Im}\,{\mathbb C}_5|^2\big) \,, ~~~
\end{align}
where \,$\alpha_{\rm e}^{}=e^2/(4\pi)=1/137$.\,
With \,$\sqrt2\,K_S=K^0-\,\overline{\!K}{}^0$,\, the amplitude for \,$K_S\to\gamma\overline\gamma$\,
and its rate are obtainable from Eqs.\,\,(\ref{MK2gg'}) and (\ref{GK2gg'}), respectively,
by making the replacements \,${\rm Re}\,{\mathbb C}\to-i{\rm Im}\,{\mathbb C}$\, and
\,${\rm Im}\,{\mathbb C}_5\to i{\rm Re}\,{\mathbb C}_5$.\,

Similarly, for \,$K\to\pi\gamma\overline\gamma$\, we find
\begin{align}
\big\langle\pi^0\gamma\big|\overline d\sigma_{\alpha\omega}^{}s\big|\,\overline{\!K}{}^0\big\rangle
& \,=\, \big\langle\pi^0\gamma\big|\overline s\sigma_{\alpha\omega}^{}d \big|K^0\big\rangle
\,=\, \frac{i\sqrt2\, a_T' e}{3} \big( \varepsilon_\omega^* {\texttt k}_\alpha^{}
- \varepsilon_\alpha^* {\texttt k}_\omega^{} \big) \,,
\nonumber \\
\big\langle\pi^0\gamma\big|\overline d\sigma_{\alpha\omega}^{}\gamma_5^{}s
\big|\,\overline{\!K}{}^0\big\rangle & \,=\, \big\langle\pi^0\gamma\big|\overline s
\sigma_{\alpha\omega}^{}\gamma_5^{}d \big|K^0\big\rangle
\,=\, \frac{\sqrt2\,a_T'e}{3}\, \epsilon_{\alpha\omega\mu\nu}^{}\,
\varepsilon^{*\nu} {\texttt k}^\mu \,, &
\nonumber \\ \label{<Km2pg'>}
\big\langle\pi^-\gamma\big|\overline d\sigma_{\alpha\omega}^{}s\big|K^-\big\rangle & \,=\,
2i a_T^{} e\, \big[ \varepsilon_\alpha^* \big(p_K^{}-p_\pi^{}\big){}_\omega^{}
- \varepsilon_\omega^* \big(p_K^{}-p_\pi^{}\big){}_\alpha^{} \big]
+ \frac{2i a_T' e}{3} \big( \varepsilon_\alpha^* {\texttt k}_\omega^{}
- \varepsilon_\omega^* {\texttt k}_\alpha^{} \big)  \,,
\nonumber \\
\big\langle\pi^-\gamma\big|\overline d\sigma_{\alpha\omega}^{}\gamma_5^{} s\big|K^-\big\rangle
& \,=\, 2 e\, \epsilon_{\alpha\omega\mu\nu}^{}\, \varepsilon^{*\mu} \bigg[ a_T^{}\,
\big(p_K^\nu-p_\pi^\nu\big) + \frac{a_T'}{3}\, {\texttt k}^\nu \bigg] \,,
\end{align}
where $p_K^{}$ and $p_\pi^{}$ denote the momenta of the kaon and pion, respectively.
From these, we derive the amplitudes for the $K_L$ and $K^-$ modes to be
\begin{align} \label{MK2pgg'}
{\cal M}_{K_L\to\pi^0\gamma\overline\gamma}^{} & \,=\, \frac{4 a_T' e}{3} \bigl[
-\big(g^{\mu\nu} g^{\varrho\varsigma}-g^{\mu\varsigma} g^{\nu\varrho}\big)\, {\rm Re}\,{\mathbb C}
+ \epsilon^{\mu\nu\varrho\varsigma}\, {\rm Im}\,{\mathbb C}_5^{} \big]
\varepsilon_\mu^* \bar\varepsilon_\nu^*\, {\texttt k}_\varrho^{} \bar q_\varsigma^{} \,,
\nonumber \\
{\cal M}_{K^-\to\pi^-\gamma\overline\gamma}^{} & \,=\, 4\bigg(a_T^{}+\frac{a_T'}{3}\bigg) e
\big[ \big( g^{\mu\nu} g^{\varrho\varsigma} - g^{\mu\varsigma} g^{\nu\varrho} \big) {\mathbb C}
+ i \epsilon^{\mu\nu\varrho\varsigma}\, {\mathbb C}_5^{} \big]
\varepsilon_\mu^* \bar\varepsilon_\nu^*\, {\texttt k}_\varrho^{} \bar q_\varsigma^{} \,.
\end{align}
They translate into the differential rates
\begin{align} \label{G'K2pgg'}
\frac{d\Gamma_{K_L\to\pi^0\gamma\overline\gamma}^{}}{d\textsl{\texttt s}_{\gamma\bar\gamma}} & \,=\,
\frac{\alpha_{\rm e}^{}\,a_T^{\prime2}\,\textsl{\texttt s}_{\gamma\bar\gamma}^2}{72\pi^2m_{K^0}^3}\,
{\cal K}^{\frac{1}{2}}\big(m_{K^0}^2,m_{\pi^0}^2,\textsl{\texttt s}_{\gamma\bar\gamma}\big)\,
\big(|{\rm Re}\,{\mathbb C}|^2+|{\rm Im}\,{\mathbb C}_5|^2\big) \,,
\nonumber \\
\frac{d\Gamma_{K^-\to\pi^-\gamma\overline\gamma}^{}}{d\textsl{\texttt s}_{\gamma\bar\gamma}} & \,=\, \frac{\alpha_{\rm e}^{} \big(3a_T^{}+a_T'\big)\raisebox{1pt}{$^2$}
\textsl{\texttt s}_{\gamma\bar\gamma}^2}{72 \pi^2 m_{K^-}^3}\,
{\cal K}^{\frac{1}{2}}\big(m_{K^-}^2,m_{\pi^-}^2,\textsl{\texttt s}_{\gamma\bar\gamma}\big)\,
\big(|{\mathbb C}|^2+|{\mathbb C}_5|^2\big) \,,
\end{align}
where $\textsl{\texttt s}_{\gamma\bar\gamma}$ stands for the invariant mass squared of
the $\gamma\overline\gamma$ pair and \,${\cal K}(x,y,z) = (x-y-z)^2-4y z$.\,
To get each of the corresponding decay rates, the integration range is
\,$0\le\textsl{\texttt s}_{\gamma\bar\gamma}\le(m_K-m_\pi)^2$.\,
As in the \,$K\to\gamma\overline\gamma$\, case, the amplitude for
\,$K_S\to\pi^0\gamma\overline\gamma$\, and its differential rate have the same expressions as
their $K_L$ counterparts in Eqs.\,\,(\ref{MK2pgg'}) and (\ref{G'K2pgg'}), respectively,
except that \,Re$\,{\mathbb C}$ (Im$\,{\mathbb C}_5$) is changed to
\,$-i{\rm Im}\,{\mathbb C}$ ($i{\rm Re}\,{\mathbb C}_5$).\,

From the $a_T^{}$ terms in Eq.\,(\ref{qsq}), we can additionally extract mesonic matrix elements
pertaining to processes induced by ${\cal L}_{ds\bar\gamma}$ without the ordinary photon.
Particularly, for the three-body channels \,$K\to\pi\pi'\overline\gamma$\, we obtain
\begin{align} \label{<Kbz2pipi'>}
\big\langle\pi^+(p)\,\pi^-(\bar p)\big|\overline d\sigma_{\alpha\omega}^{}s
\big|\,\overline{\!K}{}^0\big\rangle
& \,=\; \frac{i\sqrt2\,a_T^{}}{f}\, \epsilon_{\alpha\omega\mu\nu}^{}
\big(2\bar p^\mu+\bar q^\mu\big) p^\nu \,,
\nonumber \\
\big\langle\pi^+(p)\,\pi^-(\bar p)\big|\overline d\sigma_{\alpha\omega}^{}\gamma_5^{}s
\big|\,\overline{\!K}{}^0\big\rangle
& \,=\; \frac{\sqrt2\,a_T^{}}{f} \big[ p_\alpha^{} (2\bar p+\bar q)_\omega^{}
- p_\omega^{} (2\bar p+\bar q)_\alpha^{} \big] \,,
\end{align}
\begin{align}
\big\langle\pi^+(p)\,\pi^-(\bar p)\big|\overline s\sigma_{\alpha\omega}^{}d\big|K^0\big\rangle
& \,=\, \frac{i\sqrt2\,a_T^{}}{f}\, \epsilon_{\alpha\omega\mu\nu}^{}\,
\bar p^\mu \big(2p^\nu+\bar q^\nu\big) \,,
\nonumber \\ \label{<Kz2pipi'>}
\big\langle\pi^+(p)\,\pi^-(\bar p)\big|\overline s\sigma_{\alpha\omega}^{}\gamma_5^{}d
\big|K^0\big\rangle
& \,=\; \frac{\sqrt2\,a_T^{}}{f} \big[ (2p+\bar q)_\alpha^{}\bar p_\omega^{}
- (2p+\bar q)_\omega^{}\bar p_\alpha^{} \big] \,,
\end{align}
\begin{align}
\langle\pi^0(p)\, \pi^-(\bar p)|\overline d\sigma_{\alpha\omega}^{}s|K^-\rangle
& \,=\; \frac{i a_T^{}}{f}\, \epsilon_{\alpha\omega\mu\nu}^{}
\big[ 4\bar p^\mu p^\nu + \big(\bar p^\mu-p^\mu\big)\bar q^\nu \big] \,,
\nonumber \\ \label{<Km2pipi'>}
\langle\pi^0(p)\, \pi^-(\bar p)|\overline d\sigma_{\alpha\omega}^{}\gamma_5^{}s|K^-\rangle
& \,=\; \frac{a_T^{}}{f} \big[ 4 p_\alpha^{}\bar p_\omega^{} - 4 p_\omega\bar p_\alpha^{}
+ (p-\bar p)_\alpha^{} \bar q_\omega^{} - (p-\bar p)_\omega^{}\bar q_\alpha^{} \big] \,,
\end{align}
where we have applied the relation \,$p_K^{}=p+\bar p+\bar q$.\,
These lead to the $K_L$ and $K^-$ decay amplitudes, which can be written as
\begin{align} \label{MK2ppg'}
{\cal M}_{K_L\to\pi^+\pi^-\overline\gamma}^{} & \;=\; \frac{8a_T^{}}{f} \Big[
\epsilon_{\alpha\omega\mu\nu}^{}\, \bar\varepsilon^{\alpha*} p_-^\omega
p_+^\mu \bar q^\nu\, {\rm Re}\,{\mathbb C}
+ \big( p_+^\mu p_-^\nu - p_+^\nu p_-^\mu\big) \bar\varepsilon_\mu^*\bar q_\nu^{}\,
{\rm Im}\,{\mathbb C}_5^{} \Big] \,,
\nonumber \\
{\cal M}_{K^-\to\pi^-\pi^0\overline\gamma}^{} & \;=\; \frac{8 a_T^{}}{f} \Big[
\epsilon_{\alpha\omega\mu\nu}^{}\, \bar\varepsilon^{\alpha*} p_-^\omega p_0^\mu \bar q^\nu\,
{\mathbb C} + i \big(p_-^\mu p_0^\nu - p_-^\nu p_0^\mu\big) \bar\varepsilon_\mu^* \bar q_\nu^{}\,
{\mathbb C}_5 \Big] \,,
\end{align}
where $p_{+,-,0}$ represent the momenta of $\pi^{+,-,0}$, respectively.\footnote{Although
$\langle\pi^0\pi^0|\overline d\sigma_{\alpha\omega}^{}(1,\gamma_5^{})s|\,\overline{\!K}{}^0\rangle$
and $\langle\pi^0\pi^0|\overline s\sigma_{\alpha\omega}^{}(1,\gamma_5^{})d|K^0\rangle$ from
Eq.\,(\ref{qsq}) are not zero, their contributions to the \,$K_{L,S}\to\pi^0\pi^0\overline\gamma$\,
amplitudes vanish.
This is consistent with angular momentum conservation and gauge invariance (Bose symmetry)
forbidding the pion pair in these decays from having an angular momentum
\,$J_{\pi\pi}=0\,(1)$,\, similarly to the \,$K_{L,S}\to\pi^0\pi^0\gamma$\, case with the ordinary
photon~\cite{Lee:1966hp}.
We can therefore neglect \,$K_{L,S}\to\pi^0\pi^0\overline\gamma$,\, which are chirally suppressed
compared to the \,$K\to\pi\pi'\overline\gamma$\, modes we consider in Eq.\,(\ref{MK2ppg'}).}
We then arrive at the differential rates
\begin{align} \label{G'K2ppg'}
\frac{d\Gamma_{K_L\to\pi^+\pi^-\overline\gamma}^{}}{d\hat s} & \;=\;
\frac{a_T^2 \big(m_{K^0}^2-\hat s\big)\raisebox{1pt}{$^3$}}{96\pi^3 f^2 m_{K^0}^3 \sqrt{\hat s}}
\big(\hat s-4m_{\pi^-}^2\big)\raisebox{1pt}{$^{3/2}$}
\big[({\rm Re}\,{\mathbb C})^2+({\rm Im}\,{\mathbb C}_5)^2\big] \,,
\\
\frac{d\Gamma_{K^-\to\pi^-\pi^0\overline\gamma}^{\vphantom{|^|}}}{d\hat s} & \;=\;
\frac{a_T^2 \big(m_{K^-}^2-\hat s\big)\raisebox{1pt}{$^3$}}{96\pi^3 f^2 m_{K^-}^3 \hat s^2}\,
{\cal K}^{\frac{3}{2}}\big(m_{\pi^-}^2,m_{\pi^0}^2,\hat s\big)\,
\big(|\mbox{\small$\mathbb C$}|^2+|\mbox{\small$\mathbb C$}_5|^2\big) \,,
\end{align}
where $\hat s$ designates the invariant mass squared of the pion pair.
They are to be integrated over \,$(m_\pi+m_{\pi'})^2\le\hat s\le m_K^2$\, to yield the decay rates.
Like before, $d\Gamma_{K_S\to\pi^+\pi^-\overline\gamma}/d\hat s$ has the same formula as that in
Eq.\,(\ref{G'K2ppg'}), but with \,Re$\,{\mathbb C}$ (Im$\,{\mathbb C}_5$) replaced with
\,${\rm Im}\,{\mathbb C}$ (${\rm Re}\,{\mathbb C}_5$).

We remark that in Eqs.\,\,(\ref{<g|dsg5s|K>}), (\ref{<Km2pg'>}), and
(\ref{<Kbz2pipi'>})-(\ref{<Km2pipi'>})
each matrix element of $\overline d \sigma_{\alpha\omega}s$ and its
\,$\overline d\sigma_{\alpha\omega}\gamma_5^{}s$\, counterpart are related due to the identity
\,$2i\sigma_{\alpha\omega}\gamma_5^{}=\epsilon_{\alpha\omega\mu\nu}\sigma^{\mu\nu}$\,
for \,$\epsilon_{0123}=1$.\,
Furthermore, the amplitudes in Eqs.\,\,(\ref{MK2gg'}), (\ref{MK2pgg'}), and (\ref{MK2ppg'})
respect electromagnetic and U(1)$_D$ gauge invariance.

\section{Kaon decay predictions\label{predict}}

From the results of the preceding section for the (differential) rates of the kaon decays of
interest, we can evaluate their branching fractions in terms of the coefficients $\mathbb C$
and ${\mathbb C}_5$.
For the input parameters, we employ \,$f=f_\pi^{}=92.07(85)$\,\,MeV\, and the measured kaon
lifetimes and meson masses from Ref.\,\cite{Tanabashi:2018oca}, as well as the lattice QCD
estimates \,$a_T^{}=0.658(23)$/GeV \cite{Baum:2011rm} and
\,$a_T'=3.3(1.1)$/GeV \cite{Kamenik:2011vy,Buividovich:2009bh} at a renormalization scale of 2\,GeV.
Thus, with their central values we get
\begin{align} \label{BK2gg'}
{\cal B}(K_L\to\gamma\overline\gamma) & \,=\, 5.74\times10^{12}\,
\big[({\rm Re}\,{\mathbb C})^2+({\rm Im}\,{\mathbb C}_5)^2\big] \rm\,GeV^2 \,,
\nonumber \\
{\cal B}(K_S\to\gamma\overline\gamma) & \,=\, 1.00\times10^{10}\,
\big[({\rm Im}\,{\mathbb C})^2+({\rm Re}\,{\mathbb C}_5)^2\big] \rm\,GeV^2 \,,
\\
{\cal B}(K_L\to\pi^0\gamma\overline\gamma)^{\vphantom{\int_\int^\int}} & \,=\, 4.95\times10^9\,
\big[({\rm Re}\,{\mathbb C})^2+({\rm Im}\,{\mathbb C}_5)^2\big] \rm\,GeV^2 \,,
\nonumber \\
{\cal B}(K_S\to\pi^0\gamma\overline\gamma) & \,=\, 8.67\times10^6\,
\big[({\rm Im}\,{\mathbb C})^2+({\rm Re}\,{\mathbb C}_5)^2\big] \rm\,GeV^2 \,,
\nonumber \\
{\cal B}(K^-\to\pi^-\gamma\overline\gamma)^{\vphantom{|_|^|}} & \,=\, 2.67\times10^9\,
\big(|{\mathbb C}|^2+|{\mathbb C}_5|^2\big) \rm\,GeV^2 \,,
\\ \label{BK2ppg'}
{\cal B}(K_L\to\pi^+\pi^-\overline\gamma)^{\vphantom{\int_\int^\int}} & \,=\, 4.67\times10^{10}\,
\big[({\rm Re}\,{\mathbb C})^2+({\rm Im}\,{\mathbb C}_5)^2\big] \rm\,GeV^2 \,,
\nonumber \\
{\cal B}(K_S\to\pi^+\pi^-\overline\gamma) & \,=\, 8.18\times10^7\,
\big[({\rm Im}\,{\mathbb C})^2+({\rm Re}\,{\mathbb C}_5)^2\big] \rm\,GeV^2 \,,
\nonumber \\
{\cal B}(K^-\to\pi^-\pi^0\overline\gamma)^{\vphantom{|_|^|}} & \,=\, 1.12\times10^{10}\,
\big(|{\mathbb C}|^2+|{\mathbb C}_5|^2\big) \rm\,GeV^2 \,. ~~~
\end{align}
Clearly, the predictions for their upper values would depend on how large $\mathbb C$ and
${\mathbb C}_5$ might be, subject to the pertinent constraints.

The allowed ranges of these coefficients have recently been explored in the contexts of a couple
of new-physics models in Refs.\,\,\cite{Fabbrichesi:2017vma,Fabbrichesi:2019bmo}.
Therein it was pointed out that the most relevant restrictions on the coefficients in these NP
scenarios were from the data on kaon mixing, which receives loop contributions involving the same
new particles that participate in the loop diagrams responsible for the $ds\overline\gamma$
couplings.
Subsequently, it was shown in Ref.\,\cite{Su:2019ipw} that these interactions also gave rise to
the FCNC decays of hyperons into a lighter baryon plus $\overline\gamma$ emitted invisibly and
that the less restrained of the models could saturate the limits on the couplings inferred from
the existing data on hyperon decays~\cite{Tanabashi:2018oca}.
This implies that the current hyperon data can already translate into model-independent
restrictions on the $ds\overline\gamma$ interactions.
The extracted bounds on $\mathbb C$ and ${\mathbb C}_5$ can then be used to estimate
the maximal values of the kaon branching fractions in Eqs.\,\,(\ref{BK2gg'})-(\ref{BK2ppg'}).

To discuss the impact of the hyperon data more quantitatively, we reproduce here the branching
fractions of the aforementioned FCNC hyperon modes calculated in Ref.\,\cite{Su:2019ipw} and
expressed in terms of $\mathbb C$ and ${\mathbb C}_5$:
\begin{align} \label{B2B'g'}
{\cal B}(\Lambda\to n\overline\gamma) & \,=\, 2.75\times10^{12}
\big(|{\mathbb C}|^2+|{\mathbb C}_5|^2\big) \rm\,GeV^2 \,,
\nonumber \\
{\cal B}(\Sigma^+\to p\overline\gamma) & \,=\, 1.54\times10^{11}
\big(|{\mathbb C}|^2+|{\mathbb C}_5|^2\big) \rm\,GeV^2 \,,
\nonumber \\
{\cal B}(\Xi^0\to\Lambda\overline\gamma,\Sigma^0\overline\gamma) & \,=\, 1.61\times10^{12}
\big(|{\mathbb C}|^2+|{\mathbb C}_5|^2\big) \rm\,GeV^2 \,,
\nonumber \\
{\cal B}(\Xi^-\to\Sigma^-\overline\gamma) & \,=\, 1.32\times10^{12}
\big(|{\mathbb C}|^2+|{\mathbb C}_5|^2\big) \rm\,GeV^2 \,,
\nonumber \\
{\cal B}(\Omega^-\to\Xi^-\overline\gamma) & \,=\, 5.18\times10^{12}\,
\big(|{\mathbb C}|^2+|{\mathbb C}_5|^2\big) \rm\,GeV^2 \,. ~~~
\end{align}
These transitions, if occur, would be among the yet-unobserved decays of the hyperons.
The branching fractions of the latter have approximate maxima which we can determine
indirectly from the data on the observed channels quoted by the Particle Data
Group~\cite{Tanabashi:2018oca}.
To do so, for each of the parent hyperons, we subtract from unity the sum of the PDG
branching-fraction numbers with their errors (increased to 2 sigmas) combined in
quadrature.
We have collected the results in the third column of Table\,\,\ref{ulbf}, where the second
column displays the sums of the branching-fraction values.\footnote{In obtaining these
entries, if the PDG numbers have asymmetric errors, the lower ones are selected.}

\begin{table}[b] \bigskip
\begin{tabular}{|c||c|c|} \hline ~Hadron~ &
$\begin{array}{c}\mbox{Branching-fraction sum}\\\mbox{of observed modes}\end{array}$ &
$\begin{array}{c}\mbox{Upper limit on total branching}\\\mbox{fraction of yet unobserved modes}
\end{array}$
\\ \hline \hline
$\Lambda$ & $1.0006\pm0.0071$  & $1.4\times10^{-2}\vphantom{|_|^{|^|}}$
\\
$\Sigma^+$ & $1.0005\pm0.0042$ & $8.0\times10^{-3}\vphantom{|_|^|}$
\\
$\Xi^0$ & $1.00000\pm0.00017$  & $3.4\times10^{-4}\vphantom{|_|^|}$
\\
$\Xi^-$ & $1.00000\pm0.00042$  & $8.3\times10^{-4}\vphantom{|_|^|}$
\\
$\Omega^-$ & $1.006\pm0.011$   & $1.6\times10^{-2}\vphantom{|_|^|}$
\\ \hline
$K_L$ & $1.0044\pm0.0018$      & $1.8\times10^{-3}\vphantom{|_|^{|^|}}$
\\
$K_S$ & $1.00191\pm0.00071$    & $7.1\times10^{-4}\vphantom{|_|^|}$
\\ \hline
\end{tabular}
\caption{The second column exhibits the sums of branching fractions of all the observed
decays \cite{Tanabashi:2018oca} of the $\Lambda$, $\Sigma^+$, $\Xi^0$, $\Xi^-$, and $\Omega^-$
hyperons and of the $K_L$ and $K_S$ mesons.
The last column contains the upper limits on the branching fractions of yet-unobserved decays
of these hadrons deduced from the numbers in the second column, as explained in
the text.\label{ulbf}}
\end{table}

Comparing the hyperon entries in the last column of this table with Eq.\,(\ref{B2B'g'}), we see
that the $\Xi^0$ bound is the most stringent and leads to
\begin{align}
|{\mathbb C}|^2+|{\mathbb C}_5|^2 & \,<\, \frac{2.1\times10^{-16}}{\rm GeV^2} \,. ~~~
\end{align}
Combining this with Eqs.\,\,(\ref{BK2gg'})-(\ref{BK2ppg'}) and assuming that for the $K_L$ ($K_S$)
cases \,Im\,$\mathbb C=\mathbb C_5=0$ ($\mathbb C={\rm Im}\,\mathbb C_5=0$), we then find
\begin{align} \label{BK}
{\cal B}(K_L\to\gamma\overline\gamma) & \,<\, 1.2\times10^{-3} \,, &
{\cal B}(K_S\to\gamma\overline\gamma) & \,<\, 2.1\times10^{-6} \,,
\nonumber \\
{\cal B}(K_L\to\pi^0\gamma\overline\gamma) & \,<\, 1.0\times10^{-6} \,, &
{\cal B}(K_S\to\pi^0\gamma\overline\gamma) & \,<\, 1.8\times10^{-9} \,,
\nonumber \\
{\cal B}(K_L\to\pi^+\pi^-\overline\gamma) & \,<\, 9.8\times10^{-6} \,, &
{\cal B}(K_S\to\pi^+\pi^-\overline\gamma) & \,<\, 1.7\times10^{-8} \,,
\nonumber \\
{\cal B}(K^-\to\pi^-\gamma\overline\gamma) & \,<\, 5.6\times10^{-7} \,, &
{\cal B}(K^-\to\pi^-\pi^0\overline\gamma) & \,<\, 2.4\times10^{-6} \,. &
\end{align}
It is worth noting that the numbers in the last line for the $K^-$ decays are equal to
their $K^+$ counterparts.
Furthermore, the predictions in Eq.\,(\ref{BK}) for the modes with an ordinary photon have
uncertainties of up to about 70\% because their rates depend on $a_T'$ which has an error of
order 30\%.

The second column of Table\,\,\ref{ulbf} also lists the sums of the branching fractions
of the observed $K_{L,S}$ decay channels.
Since the central values of these numbers exceed unity by more than 2 sigmas, we may demand
that the upper limits on the branching fractions of yet-unobserved $K_{L,S}$ modes
be less than the errors shown in the second column.
Evidently, these requirements, which are quoted in the last two rows of the third column of
the table, are satisfied by the respective $K_{L,S}$ predictions in Eq.\,(\ref{BK}).

\section{Conclusions\label{concl}}

To date there have been numerous dedicated hunts for the massive dark photon, but they still
have come up empty.
If the dark photon exists and turns out to be massless, it would have eluded those quests
for the massive one.
Therefore, it is essential that future attempts to look for dark photons accommodate
the possibility that they are massless, in which case they may have nonnegligible FCNC
interactions with SM fermions via higher-dimensional operators.

In this study, we have entertained the latter scenario, specifically that in which the massless
dark photon has dipole-type flavor-changing couplings to the $d$ and $s$ quarks.
Concentrating on the implications for the kaon sector, and taking into account indirect
model-independent constraints on the $ds\overline\gamma$ interactions inferred from the available
hyperon data, we examine especially \,$K_L\to\gamma\overline\gamma$\, and
\,$K_L\to\pi^0\gamma\overline\gamma$,\, both of which have an ordinary photon in the final states,
and demonstrate that their rates can reach levels which are potentially testable by KOTO.
Moreover, \,$K^+\to\pi^+\gamma\overline\gamma$\, and \,$K^+\to\pi^+\pi^0\overline\gamma$\, can
have rates which might be big enough to be accessible by NA62.
We have previously proposed that the corresponding hyperon decays with missing energy could be
probed by BESIII.
It follows that one or more of these presently running experiments may soon discover the massless
dark photon or, if not, come up with improved restraints on the $ds\overline\gamma$ interactions.
In any case, the results of this analysis will hopefully help stimulate efforts to seek massless
dark photons in ongoing and near-future kaon and hyperon measurements.

\acknowledgements

We would like to thank Yu-Chen Tung for experimental information which motivated this research.
It was supported in part by the MOST (Grant No. MOST 106-2112-M-002-003-MY3).

\end{document}